\documentclass{jetplFRK} 
\twocolumn 

\newcommand{\version}{v4}

\lat

\title{
       Neutrino oscillations from the splitting of Fermi points
       }

\rtitle{Neutrino oscillations from the splitting of Fermi points}

\sodtitle{Neutrino oscillations from the splitting of Fermi points}

\author{F.\,R.\,Klinkhamer\,\thanks{e-mail: 
        frans.klinkhamer@physik.uni-karlsruhe.de}}

\rauthor{F.\,R.\,Klinkhamer}

\sodauthor{Klinkhamer}

\address{
         Institut f\"{u}r Theoretische Physik, Universit\"{a}t Karlsruhe (TH),
         76128 Karlsruhe, Germany}

\dates{March 30, 2004; preprint KA-TP-02-2004, hep-ph/0403285 
       (\version)}{*}      

\abstract{As was shown previously, oscillations of massless neutrinos may
be due to the splitting of multiply degenerate Fermi points. 
In this Letter, we give the details and propose a
three-flavor model of Fermi point splittings and neutrino 
mixings with only two free parameters.
The model may explain recent experimental results from the
K2K and KamLAND collaborations.
There is also rough agreement with the data on atmospheric 
neutrinos (SuperK) and solar neutrinos (SNO), but further analysis is 
required. Most importantly, the \emph{Ansatz} allows for relatively strong 
T--violating (CP--nonconserving) effects in the neutrino sector.}


\begin{document}

\maketitle

\section{Introduction}
\label{sec:Introduction}    

Neutrino oscillations are commonly associated with neutrino-mass
differences; see, e.g., Refs.~[1--3]
for three reviews. But, the different
propagation states might also be distinguished 
by some other characteristic. Two examples discussed in the 
literature are connected with violations of the equivalence
principle \cite{Gasperini,Halprin}  
and Lorentz invariance \cite{ColemanGlashow}.

Lorentz noninvariance and CPT violation as emergent phenomena in
a fermionic quantum vacuum have been discussed recently by
Volovik and the present author \cite{KlinkhamerVolovik}. 
It was noted that one possible consequence of the splitting of
multiply degenerate Fermi points (to be defined later) 
could be neutrino oscillations. The
question is whether or not this particular type of neutrino oscillation
is compatible with the experimental data. If so, we may have an
entirely new perspective on the neutrino sector.

The aim of this paper, then, is to provide an exploratory analysis
of the experimental data on neutrino oscillations from the
perspective suggested in Ref.~\cite{KlinkhamerVolovik}. 
In order to stress the difference with mass oscillations,  
we keep an eye open to the possibility that the experimental data could,
after all, be compatible with 
relatively strong  T (and CP?) violation in the neutrino sector. 

The outline of this Letter is as follows. 
In Sec.~2, 
we discuss the case of two-flavor oscillations for massless
left-handed neutrinos with Fermi point splitting.
(A Fermi point is a point in three-momentum space at which the energy
spectrum of the fermion considered has a zero.)
In Sec.~3, 
we propose a simple three-flavor model of Fermi point splittings and neutrino
mixings, which allows for strong T violation. The model has 
two free parameters, an energy scale $B_0$ and a phase $\delta$, together with 
particular fixed values (equal or close to $\pi/4$) for the three mixing angles.
In Sec.~4, 
we give the resulting expressions for the oscillation probabilities 
among the three flavors. 
In Sec.~5, 
we compare the results of the model with the experimental data on 
neutrino oscillations. The combined data from
K2K and KamLAND (with input from SuperK) appear to favor 
T violation ($\sin\delta \ne 0$) over  time-reversal invariance 
($\sin\delta=0$), but this remains to be confirmed.
In Sec.~6, 
we present some concluding remarks.

\section{Two-flavor neutrino oscillations}
\label{sec:two-flavor-oscillations}

In the limit of vanishing Yukawa couplings,
the  Standard Model fermions are massless Weyl fermions
and have the following dispersion law 
\begin{eqnarray}
\bigl( E_{a,f}({\bf q}) \bigr)^2  &=& 
\bigl|\,c\, {\bf q} + {\bf b}^{(f)}_a \bigr|^2 \,,
\label{SMdispLaw-spacelike}
\end{eqnarray}
for three-momentum ${\bf q}$ and
with ${\bf b}^{(f)}_a =0$ for the moment.
Here, $a$ labels the sixteen types of massless left-handed 
Weyl fermions in the Standard Model 
(with a hypothetical left-handed antineutrino included) and
$f$ distinguishes the three known fermion families. 

The Weyl fermions of the original Stan\-dard Mod\-el have all 
${\bf b}^{(f)}_a$ vanishing, which makes for a 
multiply degenerate Fermi point ${\bf q}={\bf 0}$. 
[\,Fermi points (gap nodes)
${\bf q}_n$ are points in three-dimensional momentum space at
which the energy spectrum $E({\bf q})$ of the fermionic 
quasiparticle has a zero, i.e., $E({\bf q}_n)=0$.]
Nonzero pa\-ra\-meters ${\bf b}^{(f)}_a$ in the dispersion law 
(\ref{SMdispLaw-spacelike}) describe 
the splitting of this multiply degenerate Fermi point.
See Ref.~\cite{KlinkhamerVolovik} for a discussion of the physics 
that could be responsible for Fermi point splitting
and Ref.~\cite{ColemanGlashow} (and references therein)
for a general discussion of Lorentz noninvariance.

Now, consider the following pattern of spacelike splittings:
\begin{equation}
{\bf b}^{(f)}_a = Y_a \;\,{\bf b}^{(f)}\,,\;\;
{\rm for} \;\; a=1,\ldots,16,\;\; f= 1,2,3 \,,
\label{SMvecbpattern} 
\end{equation}
as given by Eq.~(5.4) of Ref.~\cite{KlinkhamerVolovik}, with a minor 
change of notation.  Given the hypercharges $Y_a$ of the Stan\-dard Mod\-el 
fermions, this pattern has only three unknowns,  
the vectors ${\bf b}^{(f)}$. The Fermi point splittings 
(\ref{SMvecbpattern}), for nonvanishing ${\bf b}^{(f)}$, violate CPT but the 
induced electromagnetic CPT--odd  Chern--Simons-like term cancels out 
exactly, consistent with the tight experimental limits.
Still, there may be other effects, for example neutrino oscillations
(as long as the neutrinos are not too much affected by the mechanism of
mass generation).

We therefore focus on massless left-handed neutrinos 
(hypercharge $Y_{\nu_L}  = -1$) with Fermi point splittings (\ref{SMvecbpattern}). 
The dispersion law for a left-handed neutrino with three-momentum ${\bf q}$ 
is then given by 
\begin{eqnarray}
\bigl( E_{\nu_L,f}({\bf q}) \bigr)^2  &=& 
\bigl|\,c\, {\bf q} - {\bf b}^{(f)} \bigr|^2 \,,
\label{DispLaw-spacelike}\end{eqnarray}
with $f=1,2,3,$ for three  neutrinos. 

In this section, we restrict our attention to oscillations
between two flavors of neutrinos
(see, e.g.,  Ref.~\cite{Kayser} for further details).  
The mixing angle between the flavor eigenstates $|A\rangle$, $|B\rangle$ 
and propagation eigenstates $|1\rangle$, $|2\rangle$ will be 
denoted by  $\theta_{\rm mix}$. These propagation states 
evolve differently as long as ${\bf b}^{(1)} \ne {\bf b}^{(2)}$ in 
the dispersion law (\ref{DispLaw-spacelike}).
 
For an initial neutrino with large enough momentum $|{\bf q}|$, 
the oscillation probability from flavor $A$ to flavor $B$ over 
a travel time $t$ (travel distance $L \sim c\,t$) is readily calculated:
\begin{eqnarray}
&&
P(A \rightarrow B) \sim
\nonumber\\[2.0mm]&&
\sin^2 \bigl(\,2\, \theta_{\rm mix}\bigr) \;
\sin^2\left(\, \frac{1}{2}\;
\mbox{\boldmath $\Delta{b}$}^{(ff')} \cdot
\widehat{\mbox{\boldmath $q$}}  
\;\; L/ \hbar\, c \right)\,,
\label{PfromDeltabvector}            
\end{eqnarray}
with $\widehat{\mbox{\boldmath $q$}} \equiv  {\bf q}/|{\bf q}|$ and
$\mbox{\boldmath $\Delta{b}$}^{(ff')}\equiv {\bf b}^{(f)} - {\bf b}^{(f')}$,
for $f=1$ and $f'=2$.
The oscillation probability (\ref{PfromDeltabvector}) is anisotropic 
and energy independent. The survival probability is given by
$P(A \rightarrow A) = 1- P(A \rightarrow B)$.
Oscillation probabilities similar to Eq.~(\ref{PfromDeltabvector})
have been discussed, for example, in Sec. III B of Ref.~\cite{ColemanGlashow}.

Next, consider the following timelike splittings of Fermi points
for the massless Standard Model fermions:
\begin{equation}
b_{0a}^{(f)} =  Y_a \; b_{0}^{(f)}\,,\;\;
{\rm for} \;\; a=1,\ldots,16,\;\; f= 1,2,3 \,,
\label{SMb0pattern}
\end{equation}
as given by Eq.~(6.5) of Ref.~\cite{KlinkhamerVolovik}.
Again, the induced electromagnetic CPT--odd  Chern--Simons-like term 
cancels out exactly. The dispersion law of a massless  
left-handed neutrino is now given by 
\begin{eqnarray}
\bigl( E_{\nu_L,f}({\bf q}) \bigr)^2  &=& 
\bigl(\,c\, |{\bf q}| - b_0^{(f)}\, \bigr)^2  \,,
\label{DispLaw-timelike}
\end{eqnarray}
with $f=1,2,3,$ for three neutrinos. 
In order to stay to the usual neutrino phenomenology
as close as possible, it is assumed in this paper 
(different from Ref.~\cite{KlinkhamerVolovik})
that $b_0^{(f)}$ in Eq.~(\ref{DispLaw-timelike}) is a CP--even parameter.
The results of Sec.~5  
are, however, independent of this assumption.

For a large enough momentum of the initial neutrino,
there is again an energy-independent two-flavor oscillation probability,
\begin{eqnarray}
&&P(A \rightarrow B) \sim
\nonumber\\[2.0mm]&&
\sin^2 \bigl(\,2\, \theta_{\rm mix}\bigr) \;
\sin^2\left(\, \frac{1}{2}\;
\Delta{b}_0^{(ff')} \; L/ \hbar\, c\right)\,,
\label{PfromDeltab0}
\end{eqnarray}
with $\Delta{b}_0^{(ff')} \equiv b_0^{(f)} - b_0^{(f')}$,
for $f=1$ and $f'=2$. The first-peak distance 
(half of the wavelength $\lambda\,$) occurs at
\begin{eqnarray}                 
&&L^{\rm \,first-peak} 
  = \pi \,\hbar c \,/\, |\Delta{b}_0^{(ff')}| \approx 
\nonumber\\[2.0mm]&&
600\:{\rm km}\; \left(\; \frac{\;10^{-12}\;{\rm eV}\;}
                          {|\Delta{b}_0^{(ff')}|}\;\right) \;\,.
\label{Loscill}
\end{eqnarray}

For completeness, we also mention neutrino-mass oscillations [1--3,8]
which are based on the Lorentz-invariant dispersion law
\begin{eqnarray}
\bigl( E_{\nu,f}({\bf q}) \bigr)^2      &=& 
c^2\, |{\bf q}|^2 +  m_f^2\, c^4  \sim  
\left(\,c\, |{\bf q}| +  \frac{m_f^2\,c^3}{2 |{\bf q}|}\,\right)^2 \!,
\label{DispLaw-mass}
\end{eqnarray}
for $|{\bf q}|  \gg m_f$ and with $f=1,2,$ for two neutrinos. 
The standard result,
\begin{eqnarray}
&&P_{\rm mass-oscill}(A \rightarrow B) \sim
\nonumber\\[2.0mm]&&
\sin^2 \bigl(\,2\, \theta_{\rm mix}\bigr) \;
\sin^2\left(\, \frac{1}{2}\;
\frac{m_{f}^2 - m_{f'}^2}{2\, E_\nu}
\; L\,c^3/ \hbar\right)\,,
\label{PfromMass}
\end{eqnarray}
has, of course, the same basic structure as Eq.~(\ref{PfromDeltab0}) but is 
now energy dependent. With $\Delta m^2 \equiv m_{f}^2 - m_{f'}^2$,
the corresponding first-peak distance is
\begin{eqnarray}
&& L^{\rm \,first-peak}_{\rm \,mass-oscill}   =  \pi \,
  \bigl(2 E_\nu\,\hbar c\bigr)\, / \,\bigl(|\Delta m^2| \,c^4\bigr)\approx 
\nonumber\\[2.0mm]&& 
600\:{\rm km}
\left( \frac{E_\nu}{{\rm GeV}} \right)  
\left( \frac{2\times 10^{-3}\;{\rm eV}^2 / c^4}{|\Delta m^2|}\,\right),
\label{Lmassoscill}
\end{eqnarray}
for energies typical of ``atmospheric neutrinos'' (see Sec.~5.1).  

\section{Three-flavor splitting and tri-maximal mixing}
\label{sec:three-flavor-splitting}

For simplicity, we consider only the timelike splittings 
(\ref{SMb0pattern}) in the rest of this Letter.
Because the neutrino oscillations are energy-independent, the analysis 
of the experimental data is
entirely different from that of the usual mass oscillations.

To illustrate this point, we choose the following regular pattern 
for the Fermi point splittings of the three left-handed neutrinos
with dispersion law (\ref{DispLaw-timelike}):
\begin{eqnarray}
b_0^{(f)} &=& f \:B_0\,,\;\; {\rm for} \;\; f= 1,2,3 \,.
\label{b0pattern}     
\end{eqnarray}
In addition, we take ``tri-maximal'' values for the
mixing angles which enter the unitary matrix $V$ between 
flavor and propagation states (see Sec.~4):  
\begin{eqnarray}
\theta_{13}  &=& \arctan \sqrt{1/2} \approx \pi/5 \,, 
\nonumber\\[1mm]
\theta_{21} &=& \theta_{32}= \arctan 1=\pi/4 \,.
\label{thetapattern}
\end{eqnarray}
This neutrino mixing matrix is parametrized as
follows \cite{Barger-etal,McKeownVogel}:
\begin{eqnarray}
V & \equiv&  
\left( \begin{array}{ccc} 
1 & 0 & 0 \\ 0 & c_{32} & s_{32} \\ 0 & -s_{32} & c_{32} 
\end{array} \right) \cdot
\left( \begin{array}{ccc} 
c_{13} & 0 & s_{13}\,e^{-i\delta} \\ 0 & 1 & 0 \\
-s_{13}\,e^{i\delta}& 0 & c_{13}
\end{array} \right)
\nonumber\\[2.0mm]&&
\cdot \left( \begin{array}{ccc} 
c_{21} & s_{21} & 0 \\ -s_{21}& c_{21} & 0 \\ 0 & 0 & 1 
\end{array} \right)   \,,
\label{Vmatrix}
\end{eqnarray}
with two Majorana phases set to zero 
and the standard notation $s_{x}$ and $c_{x}$ for 
$\sin\theta_x$ and $\cos\theta_x$.

The particular values (\ref{thetapattern}) maximize,
for given phase $\delta$, the T--violation (CP--non\-con\-ser\-va\-tion)
measure  \cite{Jarlskog}
\begin{eqnarray}
J &\equiv& \frac{1}{8}\, 
\cos\theta_{13}\,\sin 2\theta_{13}\,
            \sin2\theta_{21}\,\sin 2\theta_{32}\;\sin\delta \,.
\label{J}
\end{eqnarray}
This maximality condition on $J$ is used only as a mathematical prescription 
to select unambiguously certain ``large'' values of the mixing angles.

At this moment, we do not want to speculate on possible explanations
of the relations (\ref{b0pattern}) and (\ref{thetapattern}).
There is a certain elegance to the model, with essentially two free 
parameters ($B_0$ and $\delta$).
In contrast, the standard interpretation of the experimental results  
on neutrino oscillations  \cite{Barger-etal,McKeownVogel}
has three different neutrino masses, at least two different mixing angles,
and one undetermined phase:
\begin{eqnarray}
 m_2^2 - m_1^2  &\approx& 7\times 10^{-5}\;{\rm eV}^2/ c^4 ,
\nonumber\\[2.0mm]
|m_3^2 - m_2^2| &\approx& 2\times 10^{-3}\;{\rm eV}^2/ c^4 ,
\nonumber\\[2.0mm]
\theta_{13}  \approx 0      \,,\;\;
\theta_{21}  &\approx& \theta_{32}  \approx \pi/4
\,,\;\; \delta \in [ -\pi, \pi ] \,.
\label{MassThetapattern}
\end{eqnarray}
These values would imply that T and CP violation in the neutrino sector are 
suppressed by a small value of the mixing angle $\theta_{13}$ 
[cf. Eq.~(\ref{J})], which would not be the case for the \emph{Ansatz} 
(\ref{b0pattern})--(\ref{Vmatrix}).

\section{Three-flavor neutrino oscillations}
\label{sec:three-flavor-oscillations}

Now define three flavor states $|A\rangle, |B\rangle, |C\rangle$ 
in terms of the propagation states $|1\rangle, |2\rangle, |3\rangle$, 
which have the dispersion 
law (\ref{DispLaw-timelike}) with parameters $b_0^{(f)}$, $f=1,2,3$, 
given by the pattern (\ref{b0pattern}). In matrix form, the relation is
\begin{eqnarray}
\left( \begin{array}{c}|A\rangle\\ |B\rangle\\ |C\rangle \end{array} \right)
  &=& \; V^\star \;
\left( \begin{array}{c}|1\rangle\\ |2\rangle\\ |3\rangle  \end{array} \right)
   \,,
\label{ABC=V123}
\end{eqnarray}      
where the star indicates complex conjugation.
Here, we follow the conventions of Ref.~\cite{Barger-etal},
with the mixing matrix $V$ defined by Eq.~(\ref{Vmatrix})
for the particular values (\ref{thetapattern}). 

For a large enough momentum of the initial neutrino, the energy 
differences from Eq.~(\ref{b0pattern}) give the following
oscillation probabilities:
\begin{eqnarray}            
&&P(A \rightarrow B) = 
 (2/9)\;\, \sin^2 \left(\Delta/2\right)
\nonumber\\[1mm]
&& \times  
\left( 4 +   {\sqrt{3}}\: c_\delta   + 
\bigl( 2 + 2 {\sqrt{3}}\: c_\delta \bigr) \cos \Delta - 
       2 {\sqrt{3}}  \, s_\delta \sin \Delta \right) ,
\nonumber\\[1mm] 
&&P(A \rightarrow C) = 
 (2/9)\;\, \sin^2 \left(\Delta/2\right)
\nonumber\\[1mm]
&& \times 
\left( 4 -   {\sqrt{3}}\: c_\delta   + 
\bigl( 2 - 2 {\sqrt{3}}\: c_\delta \bigr) \cos \Delta + 
        2 {\sqrt{3}} \, s_\delta \sin \Delta \right) ,
\nonumber\\[1mm] 
&&P(A \rightarrow A) = 1-P(A \rightarrow B)-P(A \rightarrow C)\, ,                 
\nonumber  
\end{eqnarray} 
\begin{eqnarray} 
&&P(B \rightarrow C) = 
 (2/9)\;\, \sin^2 \left(\Delta/2\right)
\nonumber\\[1mm]
&& \times  
\left(     13/4 - (3/4) \,\cos 2\delta  +  2\, \cos \Delta
        -  2 {\sqrt{3}} \; s_\delta\, \sin \Delta \right) ,
\nonumber\\[1mm]
&&P(B \rightarrow A) = 
 (2/9)\;\, \sin^2 \left(\Delta/2\right) 
\nonumber\\[1mm]
&& \times  
\left( 4 +   {\sqrt{3}}\: c_\delta   + 
\bigl( 2 + 2 {\sqrt{3}}\: c_\delta \bigr)  \cos \Delta + 
        2 {\sqrt{3}} \, s_\delta \sin \Delta \right) ,
\nonumber\\[1mm]
&&P(B \rightarrow B) = 1-P(B \rightarrow C)-P(B \rightarrow A)\, ,
\nonumber   \end{eqnarray} 
\begin{eqnarray} 
&&P(C \rightarrow A) = 
 (2/9)\;\, \sin^2 \left(\Delta/2\right)
\nonumber\\[1mm]
&& \times  
\left( 4 -   {\sqrt{3}}\: c_\delta   + 
\bigl( 2 - 2 {\sqrt{3}}\: c_\delta \bigr)  \cos \Delta - 
        2 {\sqrt{3}} \, s_\delta \sin \Delta \right),
\nonumber\\[1mm] 
&&P(C \rightarrow B) = 
 (2/9)\;\, \sin^2 \left(\Delta/2\right)
\nonumber\\[1mm]
&& \times  
\left(      13/4 - (3/4)\,\cos 2\delta + 2\,  \cos \Delta
        +    2 {\sqrt{3}} \; s_\delta\, \sin \Delta \right) ,
\nonumber\\[1mm]
&&P(C \rightarrow C) = 1-P(C \rightarrow A)-P(C \rightarrow B)\, ,
\label{3-flavorProbs}   
\end{eqnarray} 
with the further definition 
\begin{eqnarray}
\Delta  &\equiv&  B_0\,t /\hbar   \sim B_0\,L /(\hbar c) 
\label{Delta}   
\end{eqnarray} 
and notation $s_{\delta}$ and $c_{\delta}$ for 
$\sin\delta$ and $\cos\delta$.
For the antiparticle probabilities replace $\delta$ by $-\delta$ 
(assuming $B_0$ to be CP--even). 
The difference of the $P(X \rightarrow Y)$ and $P(Y \rightarrow X)$
probabilities in Eq.~(\ref{3-flavorProbs}), 
for $X \ne Y$ and $s_{\delta}\,\sin\Delta \ne 0$, 
implies T violation; cf. Ref.~\cite{McKeownVogel}.

For later use, we also calculate the average probabilities
$\langle P\rangle$, defined by 
integrating $\Delta$ over the interval $[0,2\pi]$ with normalization factor
$1/(2\pi)$:
\begin{eqnarray} 
&&\Bigl(\langle P(A\rightarrow B)\rangle\, ,\,
        \langle P(A\rightarrow C)\rangle\,,\,
        \langle P(A\rightarrow A)\rangle\Bigr) =
\nonumber\\[2mm]
&&\Bigl( 1/3\,,\, 1/3\,,\, 1/3  \Bigr)\,,
\nonumber\\[2mm]
&&\Bigl(\langle P(B\rightarrow C)\rangle\,,\,
        \langle P(B\rightarrow A)\rangle\,,\,
        \langle P(B\rightarrow B)\rangle\Bigr) =
\nonumber\\[2mm]
&&\Bigl( 1/4-(\cos 2\delta)/12  \,,\, 1/3 \,,\, (5+\cos 2\delta)/12 \Bigr)\,,
\nonumber\\[2mm]
&&\Bigl(\langle P(C\rightarrow A)\rangle\,,\, 
        \langle P(C\rightarrow B)\rangle\,,\,
        \langle P(C\rightarrow C)\rangle\Bigr) = 
\nonumber\\[2mm]
&&\Bigl(1/3 \,,\, 1/4-(\cos 2\delta)/12 \,,\, (5+\cos 2\delta)/12 \Bigr)\,.
\label{Averaged3-flavorProbs}   
\end{eqnarray}
These average probabilities are equal only for $\delta = \pm\, \pi/2$.

The identification of the states $|A\rangle, |B\rangle, |C\rangle$ 
with the usual neutrinos states  
$|\nu_e\rangle$, $|\nu_\mu\rangle$, $|\nu_\tau\rangle\,$ is left to 
experiment, which, after all, observes the electrons and the muons.

\section{Comparison to experiment}
\label{sec:Comparison}

In this section, we compare the model predictions (\ref{3-flavorProbs})
with two sets of data on neutrino oscillations, one from the
SuperK and K2K experiments and 
the other from the KamLAND and SNO experiments. The LSND 
results are left out of consideration, as these have not been 
confirmed by another experiment. 
We refer to two recent reviews \cite{Barger-etal,McKeownVogel}
for further details and an extensive list of references.

\section*{5.1 \; SuperK and K2K}
\label{sec:SuperKandK2K} 

With neutrino energies in the ${\rm GeV}$ range, 
SuperK \cite{SuperK1998PRL81} discovered indirect evidence for
$\nu_\mu \rightarrow \nu_x$ oscillations starting from a
distance $L \approx 500\:{\rm km}$ (corresponding to a
zenith angle of approximately $90^{\circ}$).
The same type of neutrino oscillations has also been inferred
by K2K \cite{K2K2002} at a distance $L = 250\:{\rm km}$.
Both lengths are of the same order of magnitude as Eq.~(\ref{Loscill}).

For a more precise analysis we turn to the K2K experiment. The
crucial result is now that K2K \cite{K2K2004} does \emph{not} see 
$\nu_\mu \rightarrow \nu_e$ at an appreciable
level for the length $L = 250\:{\rm km}$ where the $\nu_\mu$ flux is 
reduced by approximately 30\,\%. The quoted numbers of neutrino events are
\begin{eqnarray}
\Bigl( N_{\nu_\mu} , N_{\nu_e} , N_{\nu_\tau}  \Bigr)
\Big|^{\,\rm K2K}_{\,L = 250\:{\rm km}} \approx 
\bigl( 56 \,,\, 1  \,,\, 23?  \bigr)\,, 
\label{K2Knumbers}
\end{eqnarray}
where the number for $N_{\nu_\tau}$ has been deduced from the expected 
number $N_{\nu_\mu} \approx 80 \pm 6$ without neutrino oscillations.

Taking the phase $\delta=\pi/4$, the probabilities calculated in 
Eq.~(\ref{3-flavorProbs}) give a ``best fit'' for
\begin{eqnarray}
&&80 \times \Bigl( P(C\rightarrow C),P(C\rightarrow A),P(C\rightarrow B)
\Bigr)\Big|^{\;\delta=\pi/4}_{\;l=0.145} \;\sim 
\nonumber\\[2.0mm]&&
\bigl( 56 \,,\, 2 \,,\, 22 \bigr)\,, 
\label{b0modelnumbers}
\end{eqnarray}
with the dimensionless length $l$ defined by
\begin{eqnarray}
2\pi\,l \equiv B_0\,L /(\hbar\, c) \sim \Delta \,.
\label{l}
\end{eqnarray}
The model numbers (\ref{b0modelnumbers}) compare well with the 
``ob\-served'' numbers (\ref{K2Knumbers}).

The comparison with the K2K experiment 
allows for the following tentative identification: 
\begin{eqnarray}
\Big(\; |A\rangle = |\nu_e\rangle \,, \;\;
      |B\rangle = |\nu_\tau\rangle\,, \;\;
      |C\rangle = |\nu_\mu\rangle \;\Big)  \Big|^{\;\delta=\pi/4}\,, 
\label{ABCidentification} 
\end{eqnarray}
at least if $\delta$ is set to $\pi/4$ 
(see Sec.~5.2  
for further discussion). With $L = 250\:{\rm km}$, we also have 
\begin{eqnarray}
B_0 &\approx& 0.145  \, (h c) /( 250\:{\rm km})
\approx 7.2 \times 10^{-13}\;{\rm eV} 
\label{B0fromK2K}
\end{eqnarray} 
and a wavelength $\lambda \approx 1700 \:{\rm km}$ ($l=1$).
The statistical error on $B_0$ is estimated to be of the order of
$10 \,\%$, as obtained by letting the $N_{\nu_\tau}$ value in 
Eq.~(\ref{K2Knumbers}) range from $17$ to $29$ and finding 
the matching probabilities in the model.

The K2K experiment has also analyzed the spectrum of the reconstructed 
energies of the $\mu$--type neutrinos. Given the 
large errors, the data points agree more or less with the shape 
expected from the Fermi-point-splitting mechanism
(box histogram in Fig.~2 of Ref.~\cite{K2K2002}).

The production rates corresponding to Eq.~(\ref{b0modelnumbers})
first have a  peak for $B$--type neutrinos at 
$l \approx 0.3$ and then a peak for $A$--type neutrinos at $l \approx 0.6$, 
with the $C$--type rate reduced to under 20\,\%
over the range $0.3 \lesssim  l \lesssim  0.7$.
For SuperK, the $C$--type ($=\mu$--type?) atmospheric neutrinos would start 
being depressed at a length $L \approx 500\:{\rm km}$ ($l \approx 0.3$) 
which is roughly what is observed at a zenith angle of $90^{\circ}$. 
With travel distances averaged over several thousand kilometers
(corresponding to large enough zenith-angle intervals), the number of 
initial $C$--type neutrinos would be reduced significantly.
According to Eq.~(\ref{Averaged3-flavorProbs}) for $\delta =\pi/4$, 
an initial $2:1$ ratio of $C$--type to $A$--type neutrinos 
would be changed as follows:
\begin{eqnarray}
&&\bigl( N_{C,\bar{C}} :  N_{A,\bar{A}} :  N_{B,\bar{B}}\bigr) =  
\bigl(120 : 60 :  0 \bigr) \rightarrow 
\nonumber\\[2.0mm]&&
\bigl( 50+20  : 40+20 :  30+20 \bigr)\,.
\end{eqnarray}
Apparently, these averaged vacuum oscillations  would keep 
the initial number of $A$--type ($=e$--type?) events 
unchanged and reduce the initial number of $C$--type ($=\mu$--type?) events 
by $40\,\%$, more or less as observed by SuperK \cite{SuperK1998PRL81}. 

Needless to say, a complete re-analysis of the SuperK data 
is required, if the neutrino energy is given by Eq.~(\ref{DispLaw-timelike})
instead of the Lorentz-invariant relation (\ref{DispLaw-mass}).
The most important task would be to establish unambiguously 
whether or not the oscillation properties depend on the neutrino 
energy.
(Fig.~4 of Ref.~\cite{SuperK1998PRL81} is not really conclusive, 
because the data points can also be fitted by a smoothed steplike 
function, which drops from a constant value 
$1$ for $L/E_\nu \lesssim 100\:{\rm km}/{\rm GeV}$ 
to a constant value 
$0.6$  for $L/E_\nu \gtrsim 400\:{\rm km}/{\rm GeV}$.)

\section*{5.2 \; KamLAND and SNO}
\label{sec:KamLANDandSNO}

With antineutrino energies in the ${\rm MeV}$ range,
KamLAND \cite{KamLAND2002} presented indirect evidence for
$\bar\nu_e \rightarrow \bar\nu_x$  oscillations  at 
a distance $L \approx 180\:{\rm km}$.
The experiment quotes the following survival  probability:
\begin{eqnarray}
&&P(\bar\nu_e \rightarrow \bar\nu_e)
\Big|^{\,\rm KamLAND}_{\,L \approx 180\:{\rm km}} =
\nonumber\\[2.0mm]&&
0.611 \pm 0.085\,{\rm (stat)} \pm 0.041\,{\rm (syst)} \,.
\label{P-KamLAND}
\end{eqnarray}
The distance $L \approx 180\:{\rm km}$ corresponds to $l \approx 0.104$,
as defined by Eq.~(\ref{l}) for the tentative energy scale (\ref{B0fromK2K}).
  From Eq.~(\ref{3-flavorProbs}) specialized to $\delta=\pi/4$, the relevant 
probability for the identification (\ref{ABCidentification}) is  
\begin{eqnarray}
P\bigl(\bar{A} \rightarrow \bar{A} \bigr)
\Big|^{\;\delta=\pi/4}_{\;l=0.104} \sim  0.74 \,, 
\label{b0modelPAA}
\end{eqnarray}
which is less than two standard deviations away from the experimental
result (\ref{P-KamLAND}); cf. Fig.~4 of Ref.~\cite{KamLAND2002}. 
Note that a $10 \,\%$ error on the value of $l$ translates into 
a  $6 \,\%$ error for the probability  (\ref{b0modelPAA}).

The KamLAND experiment has also analyzed the positron energy spectrum  
from the inverse $\beta$--decay used to detect the antineutrinos.
The spectrum is reported to be consistent at 
the $53\,\%$ C.L. with the expectations from the Fermi-point-splitting 
mechanism (upper histogram in Fig.~5 of Ref.~\cite{KamLAND2002} multiplied by
a factor $0.6$).

Considering the oscillation probabilities (\ref{3-flavorProbs})
for only two values of the phase, $\delta=0$ and $\delta=\pi/4$,
the \emph{combined} experiments of K2K, KamLAND, and SuperK 
appear to favor the nonzero value of $\delta$.
As an example of a disfavored identification
(actually one of the best for $\delta=0$), we list the following numbers:
\begin{eqnarray}
&&80 \times \Bigl( P(A\rightarrow A),P(A\rightarrow C),P(A\rightarrow B)
\Bigr)\Big|^{\;\delta=0}_{\;l=0.115} \sim
\nonumber\\[2.0mm]&& 
\bigl( 56 \,,\, 2   \,,\, 22  \bigr) \,,
\label{badNumbers1}
\\[2.0mm]&&
P\bigl(\bar{C} \rightarrow \bar{C} \bigr)
\Big|^{\;\delta=0}_{\;l=0.083} \sim  0.92 \,.
\label{badNumbers2}
\end{eqnarray}
The first set of numbers compares well with the K2K data (\ref{K2Knumbers})
but the second number is rather far from the KamLAND result (\ref{P-KamLAND}).
The SuperK results, which indicate $\nu_\mu \rightarrow \nu_x$ 
wavelengths of at least $1000\:{\rm km}$,
also help to rule out certain other $\delta=0$ identifications.

The model predictions for $\delta=-\pi/4$ and $\delta=\pm\pi/2$
have also been compared  with the experimental data  
and ``best fits'' are found with numbers similar to those of
Eqs.~(\ref{badNumbers1})--(\ref{badNumbers2}) or worse.
The $\delta = \pi/4$ identification (\ref{ABCidentification})
seems to be preferred among the cases considered, at least for 
the Fermi point splittings (\ref{b0pattern}) and mixing angles 
(\ref{thetapattern}). 
Note that the $\delta$ range can be restricted to
$[-\pi/2 , \pi/2]$, since $B$ and $C$ switch roles in the 
probabilities (\ref{3-flavorProbs}) for 
$\delta \rightarrow \delta + \pi$.

The preliminary result for the T--violating phase is then
\begin{eqnarray}
\delta &\approx& -3\pi/4  \;\; {\rm or}  \;\;  \pi/4  \,,
\label{delta}
\end{eqnarray}
with identifications 
$|A\rangle = |\nu_e\rangle \,$, $|B\rangle = |\nu_\mu\rangle\,$, 
$|C\rangle = |\nu_\tau\rangle$ for the case of $\delta = -3\pi/4$
and identifications (\ref{ABCidentification}) for $\delta = \pi/4$.
A comprehensive statistical analysis remains to be performed
in order to determine the error on these values for $\delta$.

In contrast to KamLAND, the experiments of CHOOZ \cite{CHOOZ2002} 
and Palo Verde \cite{PaloVerde2001} failed to see evidence for
$\bar\nu_e \rightarrow \bar\nu_x$ oscillations at 
$\ell \approx 1\:{\rm km}$.  This would be consistent with the 
probabilities (\ref{3-flavorProbs}),
\begin{eqnarray}
&&P(\bar{A} \rightarrow \bar{A}) = 1-{\rm O}(\Delta_\ell^2\,)\,, \;\; 
P(\bar{A} \rightarrow \bar{B}) = {\rm O}(\Delta_\ell^2\,)\,, 
\nonumber\\[2.0mm]&&
P(\bar{A} \rightarrow \bar{C})= {\rm O}(\Delta_\ell^2\,)\,, 
\label{Psmallbaseline}
\end{eqnarray}
for $|\bar{A}\rangle = |\bar\nu_e\rangle \,$ 
and with  $\Delta_\ell  \equiv B_0\,\ell /(\hbar c) 
\approx \ell/(270\:{\rm km})  \ll 1$
for the  energy scale (\ref{B0fromK2K}). Remarkably, this general 
reduction of the off-diagonal oscillation probabilities
does not require $\sin 2\theta_{13}$ to be close to zero
as would be the case for mass oscillations
(\ref{PfromMass})--(\ref{Lmassoscill})    
with $|m_1^2 - m_3^2| \approx 2\times 10^{-3}\;{\rm eV}^2 / c^4$
and $E_\nu \approx  3 \;{\rm MeV}$ \cite{CHOOZ2002,PaloVerde2001}.

As to solar neutrinos,
SNO \cite{SNO2002,SNO2003} has definitely established flavor 
oscillations, with the initial $e$--type neutrinos 
distributed over the three flavors and their flux 
reduced to approximately $30\,\%$. Vacuum oscillations of neutrinos
with Fermi point splittings (\ref{b0pattern})
and tri-maximal mixing angles (\ref{thetapattern}) have 
an average $e$--type survival probability of $1/3$, according to 
Eqs.~(\ref{Averaged3-flavorProbs}) and (\ref{ABCidentification}). 
Matter effects can be expected to play a 
role because the matter-oscillation length scale
$h c/(2\sqrt{2}\; G_F \, n_e )$ is  approximately 
$100\:{\rm km}$ in the center of the Sun \cite{McKeownVogel}, 
which is definitely less than our length scale (\ref{Loscill}).
But, in the end, matter effects may be rather unimportant if 
the vacuum mixing angles are close to $\pi/4$.

It is not clear how well our neutrinos with Fermi point 
splittings fit \emph{all} the solar neutrino data from SNO 
and the other experiments \cite{Barger-etal,McKeownVogel}. 
Obviously, a complete re-analysis of neutrino 
propagation in the Sun is required,   
if the vacuum dispersion law is given by Eq.~(\ref{DispLaw-timelike}).

There is also the possibility of further effects from small neutrino 
masses with their own matrix  structure; cf. Ref.~\cite{ColemanGlashow}. 
These small masses could affect flavor oscillations of solar neutrinos 
with relatively low energy ($E_\nu \lesssim 1\;{\rm MeV}$
for $|\Delta m^2|  \approx 10^{-6}\;{\rm eV}^2 / c^4\,$),
whereas oscillations of neutrinos with higher energy would be 
primarily determined by the Fermi point splittings (\ref{B0fromK2K}).

\section{Conclusion}
\label{sec:Conclusion}

The present Letter has shown that energy-in\-de\-pen\-dent
neutrino oscillations from the timelike splitting of Fermi 
points \cite{KlinkhamerVolovik} need  not be in flagrant 
contradiction with the experimental data [10--17].

For the sake of argument, we have considered a simple
model (\ref{b0pattern})--(\ref{Vmatrix})  with two free parameters,
the energy scale $B_0$ and the phase $\delta$.
The mixing angles of this model are fixed to the values
(\ref{thetapattern}) by the condition that they
maximize the function (\ref{J}) for a given phase $\delta$.
It turns out that the model can more or less explain the results of the
K2K and KamLAND experiments, with a fundamental energy scale $B_0$
of the order of $10^{-12}\;{\rm eV}$
and a preference for a nonzero T--violating phase,
$\,\sin^2\delta \approx  1/2\,$. 
(These numerical values are, of course, to be considered preliminary.) 
There is also rough agreement with the data on
atmospheric neutrinos (SuperK) and solar neutrinos (SNO), but 
further analysis is needed.

The tentative conclusion is that the simple \emph{Ansatz}
(\ref{b0pattern})--(\ref{thetapattern}) for the neutrino dispersion 
law (\ref{DispLaw-timelike}) and mixing matrix (\ref{Vmatrix}) 
may be compatible with experiment. 
The model considered can, of course, be perturbed by 
changes in the energy scales and mixing angles and by the
addition of small mass terms. More importantly, the \emph{Ansatz}    
suggests an entirely new structure of the neutrino sector, 
with the possibility of relatively strong T (and CP?) violation.

We end this Letter with four general remarks.
First, the spacelike splitting of Fermi points is a possibility 
not considered in detail here, as the phenomenology would certainly be 
more complicated due to the presence of
anisotropies; cf. Eq.~(\ref{PfromDeltabvector}).
Second, left-handed antineutrinos  (with hypercharge $Y_{\bar\nu_L}  =0$)
drop out for the patterns (\ref{SMvecbpattern}) and (\ref {SMb0pattern}) 
but could perhaps play a role  in further mass generation.
Third, it remains to be seen how the Fermi point splittings of 
the massless (or nearly massless) neutrinos feed into the charged-lepton 
sector; cf. Ref.~\cite{MocioiuPospelov}.
Fourth, theory and experiment need to elucidate the precise 
role, if any, of CP, T, and CPT violation in the neutrino sector.

The author thanks G.\,E. Volovik for extensive discussions
and M. Jezabek and the referees for useful comments on the manuscript.

\end{document}